\documentclass[10pt, conference]{IEEEtran}
\IEEEoverridecommandlockouts
\usepackage{cite}
\usepackage{amsmath,amssymb,amsfonts}
\usepackage{graphicx}
\usepackage{textcomp}
\usepackage{xcolor}
\usepackage{amsmath}
\usepackage{bbm}
\usepackage{svg}
\usepackage{subfigure}
\usepackage{booktabs}
\usepackage{textcomp}
\usepackage{balance}
\usepackage{algorithmic}
\usepackage{graphicx}
\usepackage{multirow}
\usepackage[utf8]{inputenc}
\usepackage[ruled]{algorithm2e}
\def\BibTeX{{\rm B\kern-.05em{\sc i\kern-.025em b}\kern-.08em
    T\kern-.1667em\lower.7ex\hbox{E}\kern-.125emX}}
    
\begin{document}

\title{Few-shot Multi-domain Knowledge Rearming for Context-aware Defence against Advanced \\Persistent Threats}

\author{
	\IEEEauthorblockN{
		Gaolei Li \IEEEauthorrefmark{1}, 
            Yuanyuan Zhao \IEEEauthorrefmark{2}
        Wenqi Wei \IEEEauthorrefmark{3},
        and Yuchen Liu \IEEEauthorrefmark{4}  
  }\\
	\IEEEauthorblockA{\IEEEauthorrefmark{1}School of Electronic Information and Electrical Engineering, Shanghai Jiao Tong University, China\\}
	\IEEEauthorblockA{\IEEEauthorrefmark{2} School of Information Science and Engineering, Hangzhou Normal University, China\\}
    \IEEEauthorblockA{\IEEEauthorrefmark{3}Computer and Information Sciences, Fordham University, USA \\}
	\IEEEauthorblockA{\IEEEauthorrefmark{4} Department of Computer Science, North Carolina State University, USA \\}
 gaolei\_li@sjtu.edu.cn, yyzhao04@163.com, wenqiwei@fordham.edu, yuchen.liu@ncsu.edu}

\maketitle

\begin{abstract}
Advanced persistent threats (APTs) have novel features such as multi-stage penetration, highly-tailored intention, and evasive tactics. APTs defense requires fusing multi-dimensional Cyber threat intelligence data to identify attack intentions and conducts efficient knowledge discovery strategies by data-driven machine learning to recognize entity relationships. However, data-driven machine learning lacks generalization ability on fresh or unknown samples, reducing the accuracy and practicality of the defense model. Besides, the private deployment of these APT defense models on heterogeneous environments and various network devices requires significant investment in context awareness (such as known attack entities, continuous network states, and current security strategies). In this paper, we propose a few-shot multi-domain knowledge rearming (FMKR) scheme for context-aware defense against APTs. By completing multiple small tasks that are generated from different network domains with meta-learning, the FMKR firstly trains a model with good discrimination and generalization ability for fresh and unknown APT attacks. In each FMKR task, both threat intelligence and local entities are fused into the support/query sets in meta-learning to identify possible attack stages. Secondly, to rearm current security strategies, an finetuning-based deployment mechanism is proposed to transfer learned knowledge into the student model, while minimizing the defense cost. Compared to multiple model replacement strategies, the FMKR provides a faster response to attack behaviors while consuming less scheduling cost. Based on the feedback from multiple real users of the Industrial Internet of Things (IIoT) over 2 months, we demonstrate that the proposed scheme can improve the defense satisfaction rate.

\end{abstract}


\section{Introduction}
Advanced Persistent Threats (APTs) are sophisticated and targeted cyber-attacks that are designed to gain unauthorized access to a system or network and remain undetected for an extended period. 
APTs typically involve a multi-stage attack process, which may include reconnaissance, initial access, escalation of privileges, lateral movement, data exfiltration, and maintaining persistence in the targeted system or network. Collecting and analyzing threat intelligence about known APT groups, their tactics, techniques, and procedures (TTPs), and their target sectors can help organizations identify potential threats and take proactive measures to mitigate them. Meanwhile, deploying endpoint security solutions, such as antivirus software, intrusion detection and prevention systems (IDPSs), and endpoint detection and response (EDR) tools, can help detect and prevent APTs from compromising endpoints. Besides, providing a well-defined incident response plan in place can help organizations quickly respond to and contain an APT attack, minimize the damage caused, and restore normal operations as soon as possible. 

\begin{figure}
\centering
\includegraphics[width=3.2in]{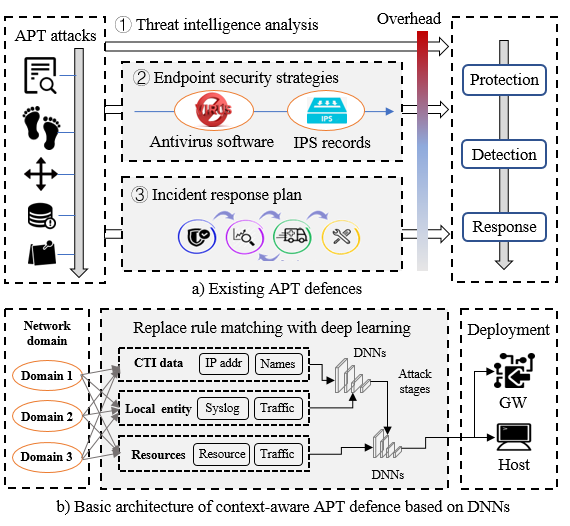}
\caption{The workflow of context-aware APT defense based on DNNs. Different from the traditional protection, detection, and response (PDR) model, the context-aware APT defense model based on DNNs focuses on precisely identifying attack stages and adaptively deploying DNNs in each network domain.}
\label{fig:CAPTD}
\end{figure}

Context-aware APT defense refers to a type of advanced threat defense system that utilizes contextual information to detect and respond to APT attacks, as illustrated in Fig \ref{fig:CAPTD}. It replaces traditional rule matching-based methods \cite{5462149,6914042} with machine learning-based methods \cite{8264962,9627773,7841721}. Context-aware APT defense is a proactive method of threat detection and response that can help organizations stay ahead of sophisticated APT attacks \cite{9951012}. This method analyzes network traffic, user behavior, and other contextual information to detect work stages of anomalies and suspicious activities that may indicate an APT attack, relying on various machine learning-based behavioral analytics \cite{singh2019comprehensive}. Once an APT entity is detected, this system will respond by isolating the affected system or network segment, blocking malicious traffic, and alerting security personnel. 

While the context-aware APT defense technologies mentioned in above are effective in mitigating the risks of APT attacks, they also have some drawbacks: 1) False positives: Some APT defense methods, such as intrusion detection systems (IDSs) and security information and event management (SIEM) platforms, may generate false positives, leading to wasted time and resources investigating benign events; 2) Cost: Implementing APT defense measures can be expensive, as it often requires specialized hardware, software, and personnel; 3) Complexity: APT defense methods can be complex to deploy, configure, and maintain, requiring significant expertise and training; 4) False negatives: APT defense methods may not detect all APT attacks, particularly those that use advanced evasion techniques or zero-day vulnerabilities. Through observation, we find that the bane of these drawbacks lies in the strong independence and poor coordination among different defense methods.

To address these problems, we propose a novel few-shot multi-domain knowledge rearming (FMKR) scheme for context-aware defense against APTs. By completing multiple small tasks that are generated from different network domains with meta-learning, the FMKR firstly trains a model with good discrimination and generalization ability for fresh and unknown APT attacks. In each FMKR task, threat intelligence and local entities are fused into the support/query sets in meta-learning to identify possible attack stages. Secondly, to improve defense efficiency, an adversarial on-demand distillation (AOD) mechanism is proposed to transfer learned knowledge into student models, which can be inexpensively deployed in each network domain. Compared to the state-of-the-art resource allocation-assisted defense deployment methods, the FSKR provides a faster response to attack behaviors while consuming less scheduling cost. Based on the feedback from multiple real users of IIoT over 2 months, we demonstrate that the proposed scheme can improve the defense satisfaction rate.


\begin{itemize}
\item A novel few-shot multi-domain knowledge rearming (FMKR) scheme is proposed for context-aware defense against advanced persistent threats. The FMKR formulates context-ware APT defense as the meta-learning task, including meta-training and meta-testing phases. With the introduction of meta-learning, the FMKR converges quickly even when the number of collected data samples corresponding to each attack stage is few and their distribution is imbalanced.  
\item To improve the reliability and adaptability of context-aware APT defense on unknown attacks, we also present a finetuning-based deployment mechanism, in which the used fine-tuning data are sampled from the target network domain in real time. Finetuned models can be deployed on various devices without the constraints of boundary security theory. 
\item Both simulations based on datasets and real observations in IIoT scenarios demonstrate the feasibility of the proposed methods.
\end{itemize}

The rest of this paper is structured as follows. In Section II, we give a comprehensive overview of context-aware APT detection and few-shot learning to highlight the necessity of our work. And then, we introduce the framework of the proposed FMKR scheme and its workflow in detail in Section III. Subsequently, we introduce experimental evaluations and real observations in Section IV. Finally, we give the concluding remarks in Section V.

\section{Related Work}
In the past few decades, cloud-based threat detection systems have made unprecedented progress. Moreover, many machine learning-based security products such as FireEye Helix, Palo Alto Networks Cortex XDR and Cisco Umbrella have played a huge role in defending against network attacks. However, as network attacks become increasingly intelligent, profit-driven, and organized, traditional security boundaries are gradually disappearing. Therefore, the adoption of zero-trust, few-shot, and adaptive security policies has become an inevitable trend in the development of the era. The purpose of these policies is to enhance the reliability and flexibility of network security, in order to better protect the network security of enterprises and users.

\subsection{Toward Context-aware Detection against APTs}
The evolution of APT detection technology has been driven by the increasing sophistication and complexity of APT attacks, as well as the need to detect and respond to these attacks in real time. The APT detection technology has gone through several stages: 1) Signature-based detection: It involved creating signatures, or patterns of known malicious code, and using them to identify and block known APT attacks \cite{1029407}. 2) Behavior-based detection: It involves monitoring for unusual or malicious behavior, such as lateral movement, privilege escalation, and data exfiltration, that may indicate an APT attack \cite{9802721}. Therein, abnormal detection focuses on identifying deviations from a baseline of normal activity by correlating large volumes of data from multiple sources, such as logs, network traffic, and endpoint behavior \cite{9632825}. 3) Threat intelligence: It involves collecting and analyzing information about potential APT threats, such as indicators of compromise and attack methods, and using this information to proactively detect and respond to APT attacks \cite{9204463}. 

Emergence of ATT$\&$CK and Kill Chain models has promoted the development of context-aware APT defence, which can help security teams better understand and assess the threat of network attacks, and provide a structured approach to collecting and analyzing threat intelligence. These models can also help security teams make better security decisions and enhance collaboration and information sharing among security teams. Context-aware APT detection involves using artificial intelligence algorithms to analyze large volumes of data in real time, identify patterns and anomalies, and respond to potential APT threats. Graph learning \cite{9802721}, semi-supervised learning \cite{9632825}, federated learning \cite{9409113,10118885}, and multi-modal learning \cite{mmham,8443370} have been studied to discover features of each APT attack stage. However, the low correlation and slow time-varying characteristics of APT attacks make the detection model have a high false alarm rate. Meanwhile, the imbalanced distribution of collected threat intelligence and local entities severely restricts the detection accuracy and generalization ability. Besides, due to the lack of real-time information about heterogeneous network environments that will be deployed, many small and medium-sized enterprises are unable to afford the cost of deploying complex APT defense strategies.

\subsection{Few-shot Learning and Knowledge Transferring}
Since most security agencies only can intercept a limited number of APT attack samples, few-shot learning, and knowledge transferring may become important enabling technologies of context-aware APT detection. Authors in \cite{9500667} firstly constructed a few-shot learning framework for SCADA systems to detect malicious intrusions, which has low computing cost so that can work on edge devices and endpoints. Literature \cite{9551702} specified that training an APT detection model on edge devices and endpoints may suffer from an imbalanced data distribution problem and designated a class balance loss to improve the few-shot learning framework. But the robustness of few-shot learning is still very weak. 
T. Ye, et al. \cite{9642512} proposed to use a Latent Dirichlet Allocation (LDA)-based pseudo samples generation algorithm to generate more attack samples to deal with the non-robustness. 
Nowadays, few-shot intrusion detection \cite{9083983} has achieved a high accuracy of up to 99.62\%, and few-shot unseen malware detection also has obtained more than 90\% accuracy \cite{9533759}. However, we find that most of existing few-shot learning methods have low practicality because 1) heterogeneous network environments can not run these few-shot learning models without software upgrading; 2) Model training process does not consider the current defense strategies so that detection results have a high delay in associating with response measures; 3) it is very hard to transfer the knowledge of APT detection model in one network domain to another domain for complex correlation of users and systems lacks self-similarity \cite{8606252, coulter2022domain,tong2022gradmfl}.


\section{Few-shot Multi-domain Knowledge Rearming for Context-aware APT Defence}
In this section, we first formulate the problems that we will resolve in this article. And then, we introduce the framework of few-shot multi-domain knowledge rearming (FMKR) for context-aware APT defense in detail.

\subsection{Proposed FMKR framework}
In a common deep learning task, a basic objective is to find a proper network parameter $\theta$ to correctly map given inputs into the corresponding labels. To obtain such a model parameter, we have to collect massive data samples and labels, being denoted as $\mathcal{D} = \{(x_1, y_1), (x_2, y_2),...,(x_i,y_i),...,(x_N, y_N)\},$ $ x_i \in \mathbb{R}^d, y_i \in \{0,1\}$ to estimate the $\theta$. Usually, the model parameter is updated using the stochastic gradient descent (SGD) algorithm, following the equation of $\theta_{t+1} \leftarrow \theta_t - \eta \nabla \mathcal{L}_{\theta}(x_i, y_i)$. With the supervised learning mode, the essence of model training is to constantly adjust the classification boundary to fit all the given data samples and labels. However, if only training samples are considered but testing samples are not considered, the obtained model parameters will be difficult to have the human-like generalization ability, so the model $f_{\theta}$ will make mistakes on $x_i + \delta$, where $\delta$ is a perturbation on $x_i$.

Moreover, in a real network environment, the distribution of training data is always imbalanced. For example, ``NT" stands for ``normal traffic", ``RN" is ``Reconnaissance", ``EF" denotes ``Establish Foothold", ``LM" represents ``Lateral Movement", and ``DE" is ``Data Exfiltration". The number of ``DE" samples is very small ($\le 10$), while the number of ``NT" samples is very large ($\ge 30000$). When we collect these samples to train the M-classification model, the obtained M-classification model $f$ can achieve high accuracy on ``NT" samples, but low accuracy on ``DE" samples. Therefore, when the network is attacked by unknown threats (only some stages are known), $f$ cannot identify whether the attack is successful because it only learns a little knowledge about ``DE".

\begin{figure}
\centering
\includegraphics[width=3.5in]{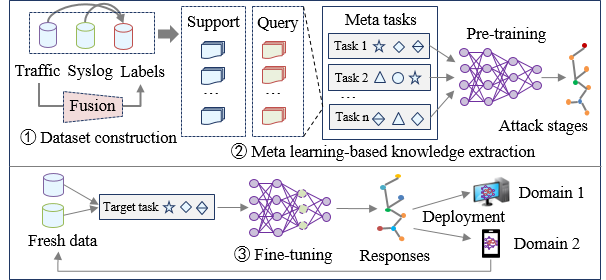}
\caption{The workflow of proposed FMKR. Both network traffic and syslog records are sampled, paired, labeled, and split into multiple meta-learning tasks. Each FMKR task contains support/query pairs, which are essential to improve the model generalization ability to unknown attack samples. To adapt to different network domain, the pre-trained model is fine-tuned with fresh samples before being deployed on edge devices. With FMKR, we can obtain a scalable, accurate, and adaptive threat detection model against multi-stage APT attacks.}
\label{MTIL}
\end{figure}
To address the above few-shot, unreliability, and data imbalance problems, We have taken three important steps: 1) Traffic and Syslog fusion, 2) Meta learning-based knowledge extraction (MKE), and 3) Finetuning-based deployment. The traffic and syslog fusion is implemented using timestamp alignment, thus the dataset $\mathcal{D}$ is expanded to $\hat{\mathcal{D}} = \{(x^t_1:x^s_1, y_1), (x^t_2:x^s_2, y_2),..., (x^t_N:x^s_N, y_N)\}$. The MKR no longer focuses on optimizing a single task but serves to transfer knowledge extracted from given tasks to a new task. In meta-learning, each task is isomorphic (that means they are all k-classification problems), but the number of training samples used for each task can be very different.  
Samples consisting of ``NT, RN, EF", ``NT, RN, LM", ``NT, EF, LM" and ``RN, EF, LM" are added to the support set, while the others are added to the query set. In this case, the ``DE" and its unknown variants can be detected. The finetuning-based deployment is proposed to improve the adaptability of the trained meta-learning model to local samples, which modifies some neural network layers on users' demands and retrain the neural networks with additional information about network states. Fig. 2 shows the entire framework of FMKR.

\subsubsection{Traffic and Syslog Fusion for Dataset Construction}
To integrate and analyze the syslog records and traffic files, we use the timestamp to align the data flow with syslog records. The limited time window is $2s$, which means if the timestamp difference between specific data flow and syslog record is within $2s$, they are remarked as the same label. 
After the above alignment processing, there are 642 syslog samples left, of which there are only two types of abnormal samples and a total of 17 samples. For APT attacks, completely relying on the timestamp to implement label fusion is inefficient for many attack behaviors are hidden and latent in the IIoT network. Therefore, all pre-processed samples is packed into a query set and a support set. 
We denote the network traffic as $x^t$ and the syslog data as $x^s$. Therefore, the paired sample can be denoted as $X = \{x^t_n; x^s_n\}, X \in \mathbb{R}^{n \times d}$, where $n$ represents the batch size, $d$ denotes the dimension of the sample vector. The label set of paired samples is denoted as $Y$, which are generated according to attack stages. 
Support/query sets are generated according to Algorithm \ref{SPG}. 

\begin{algorithm}[ht]
   \caption{Support\&query sets generating via fused data}
   \label{SPG}
\begin{algorithmic}[1]
   \STATE {\bfseries Input:} Sampled network traffic $x^t$; Sampled syslog records $x^s$; Initialize meta task $T_0$; Label set $Y$.
   \STATE {\bfseries Output:} Support set $\mathcal{D}_S$; Query set $\mathcal{D}_Q$.
   \STATE Set $i=0$, $j=0$, $i \ge j$;
    \FOR{each network traffic $x^t_i \in x^t$}
    \STATE Identify the timestamp $T^t_i$ of $x^t_i$;
        \FOR{each syslog record $x^s_j \in x^s$}
        \STATE Identify the timestamp $T^s_j$ of $x^s_j$;
        \IF{$T^s_j \le T^t_i+\theta \& T^s_j \ge T^t_i-\theta$}
        \STATE Fuse $X^f = x^t_i$ and $x^s_j$ as $x^t_i:x^s_j$;
        \ELSE
        \STATE Zero Padding $X^f= x^t_i:0$;
        \ENDIF
        \ENDFOR
    \ENDFOR
\STATE Fuse the $Y^t$ and $Y^s$ as $Y^f$
\FOR{Each label $y \in Y^f$}
\IF{$y$ == ``DE"}
\STATE Put the corresponding sample into $\mathcal{D}_Q$
\ELSE
\STATE Put the corresponding sample into $\mathcal{D}_S$
\ENDIF
\ENDFOR
\STATE Return $\mathcal{D}_S$ and $\mathcal{D}_Q$
\end{algorithmic}
\end{algorithm}

\begin{algorithm}
   \caption{Training via MKR}
   \label{MMT}
\begin{algorithmic}[1]
   \STATE {\bfseries Input:} Dataset for Task slices $D=\left\{\boldsymbol{D}_S, \boldsymbol{D}_{Q}\right\}$; Hyper-parameter $\alpha$ and $\beta$; Initialize model parameters $\theta_0$; The total training round $T$.
   \STATE {\bfseries Output:} Model parameter $\theta_T$.
   \STATE Obtain the initialized $\theta_0, \alpha, \beta$.
    \FOR{each training round $t \in T$}
    \STATE \textbf{On the cloud:}
    \STATE Parsing the received messages from distributed endpoints.
    \IF{S == True}
    \STATE Execute model replacement.
    \ENDIF
        \STATE Randomly pick a batch of tasks;
        \FOR{each task $\tau_i \in \mathcal{T}$}
        \STATE Compute $\mathcal{L}_{D_S^{\tau_k}}(\theta^t) \leftarrow \frac{1}{|D_S^{\tau_k}|} \sum_{(x, y) \in D_S^{\tau_k}} \ell(f_{\theta^t}(x), y)$;
        \STATE Update the model with centralized support set:\\
        $\theta^t\leftarrow \theta^t - \alpha \nabla \mathcal{L}_{D_S^{\tau_k}}(\theta^t)$;
        \ENDFOR 
    \STATE \textbf{On the edge/endpoint:}
    \STATE $\mathcal{L}_{D_Q^{t}}(\theta^t) \leftarrow \frac{1}{|D_Q^{t}|} \sum_{(x', y') \in D_Q^{t}} \ell(f_{\theta^t}(x'), y')$;
    \STATE Update the model again via distributed query set: \\
    $\theta_t \leftarrow \theta_t - \beta \nabla \mathcal{L}_{D_Q^t}(\theta^t)$
    \STATE Send a message to the cloud for requesting model replacement.
    \ENDFOR
\STATE Output the model parameter $\theta_T$.
\end{algorithmic}
\end{algorithm}

\subsubsection{Meta learning-based Knowledge Extraction}
To guarantee the generalization ability, we first use meta-learning as the fundamental model to achieve context-aware APT detection. And then, 
to deal with data imbalance, we introduce a class equilibrium loss, that is, configuring different costs on the normal and abnormal categories so that the calculation process of the loss function pays more attention to the categories with higher costs. The cost of each class is the ratio of the maximum number of samples in all classes to the number of samples in this class: $\lambda_c = \frac{m_{max}}{m_c}$, where $\lambda_c$ denotes the cost of $c$ class, $c \in [0, C]$; $m_{max}$ represents the maximum number of samples in all categories; $m_c$ represents the number of samples in $c$ category, and $n = \sum_c^C m_c$. Therefore, the class balance loss is defined as follows:
\begin{equation}
\mathcal{L}_D = \sum_c^C \lambda_c \mathcal{L}_c = - \sum_c^C \lambda_c log \frac{exp(x_c,y_c)}{\sum_{c=1}^C exp(x_c, y_c)}
\end{equation}
where $\mathcal{L}_c$ denotes the cross entropy loss on class $c$, and $\mathcal{L}_D$ denotes the total loss on a dataset $D$. 

In the classical meta learning algorithm \cite{pmlr-v70-finn17a}, each training round contains an outer loop and an inner loop. Combined with Eq. (1), the loss function in the inner loop that can deal with imbalanced data samples is formulated as follows: 
\begin{equation}
\mathcal{L}_{D_S^{\tau_k}}(\theta^t) \leftarrow \frac{1}{|D_S^{\tau_k}|} \sum_{(x_c, y_c) \in D_S^{\tau_k}} \mathcal{L}_c (f_{\theta^t}(x_c),y_c) 
\end{equation}
\begin{equation}
\theta^t\leftarrow \theta^t - \alpha \nabla \mathcal{L}_{D_S^{\tau_i}}(\theta^t)
\end{equation}
where $\mathcal{T} = \{\tau_k\}, k \in [1, K]$ denotes a batch of meta tasks. With the inner loop, we can obtain a pre-trained model that can learn a good initial weight to achieve fast adaptation on new tasks.

The outer loop utilizes the model pre-trained in the inner loop to compute the query loss and gradients. With query gradients, we can update model parameters again to improve the model reliability on new tasks.
\begin{equation}
\mathcal{L}_{D_Q^{t}}(\theta^t) \leftarrow \frac{1}{|D_Q^{t}|} \sum_{(x'_q, y'_q) \in D_Q^{t}} \mathcal{L} (f_{\theta^t}(x'_q), y'_q)
\end{equation}
\begin{equation}
\theta_t \leftarrow \theta_t - \beta \nabla \mathcal{L}_{D_Q^t}(\theta^t)
\end{equation}

In the MKR task, the final output vector of the neural network is the probability that the sample corresponds to each type of label (attack stage). To make the output of the neural network as close to the true label as possible, the mean square error (MSE) loss can generally be used to calculate the distance between the output vector and the true label. However, in most meta learning tasks, only the category with the highest probability is generally taken as the final prediction result, and the output vector does not need to be exactly equal to the true label. Different from the MSE loss, the cross entropy loss only focuses on the probability corresponding to the correct classification result and can establish a classification boundary with a strong generalization ability in practice. However, in the field of threat detection, the goal of model training is not only to pursue high accuracy but also to prevent false positives and false negatives. 

\subsubsection{Finetuning-based Deployment}
Although the meta learning can significantly improve the generalization ability and the class balance loss can accurate the convergence speed, we still need to enhance the accuracy of the proposed FMKR on new tasks. Thus, we propose an finetuning-based deployment mechanism, which can be implemented with three different modes: 1) remove and replace the classification layer; 2) re-initialize $N$ layers and replace the classification layer; 3) extend with $N$ more layers and replace the classification layer. In order to fix the knowledge of the pre-trained model, we adopt the third fine-tuning mode to implement on-demand model fine-tuning. Firstly, we freeze $M$ layers' model weights in the pre-trained model and then extend $N$ new layers. Secondly, we train the extended $N$ layers with local data samples. Thirdly, the classification layer is also removed and replaced according to the number of attack stages. During the fine-tuning process, we use new samples that are collected from the local network in real-time to query and update the pre-trained model.

It can be seen that meta-learning and fine-tuning have been used in many machine vision algorithms, thus \textbf{what is the novelty of the proposed FMKR framework?} Firstly, the training process of the proposed FMKR is cloud-edge synergistic. In the cloud data center, there are sufficient data samples that can act as the support set of FMKR to train a better model initialization parameter. In the edge/endpoint, an FMKR agent is responsible to fine-tune the model via distributed query sets on demand. To this end, the FMKR is the first work to integrate fine-tuning with meta-learning. Secondly, we use a model replacement trick to share the knowledge to enhance the model adaptability to unknown attacks, which further improves the model generalization ability.

\section{Experiment}

\subsection{Experimental Setup}

Industrial Internet of Things (IIoT) can provide massive data for accurate mapping between virtual space and physical entities \cite{9772337,8377998}. 
However, according to extensive investigations, IIoT also brings many emerging security challenges into traditional industrial control networks. On one hand, more remote control functions and frequent data exchanges are supported in IIoT so that new system vulnerabilities may be exploited by attackers. Furthermore, more convert intrusion paths may be discovered \cite{9431233,9520645}; On the other hand, due to the diversity and heterogeneity of IIoT devices, the methods of identifying unknown security threats must be upgraded to support artificial intelligence, which can rapidly extract the high-dimensional features of sampled networks traffic data and syslog files \cite{9802721,9321458,9403369}. Our methods are implemented on a real IIoT platform.

\begin{figure}
\centering
\includegraphics[width = 3.3in]{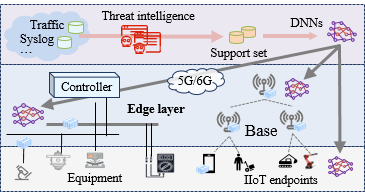}
\caption{Deploying the proposed FMKR in 5G/6G-enabled IIoT, where no existing security boundaries are assumed.}
\label{fig1}
\end{figure}

\begin{figure*}[ht]
\centering
\includegraphics[width=6.7in]{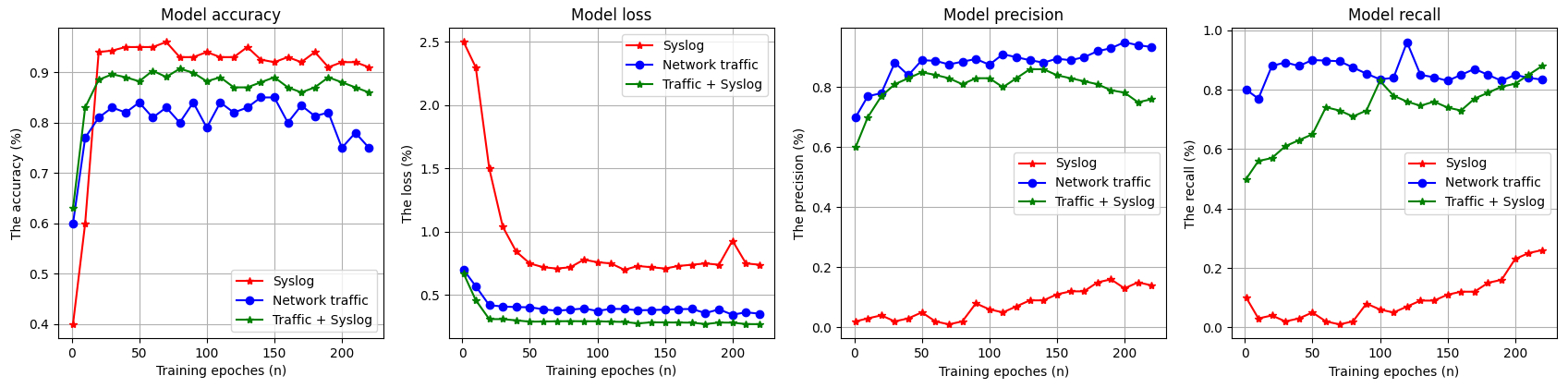}
\caption{The performance of ResNet-based FMKR method. It can be seen that fusing syslog and network traffic can achieve compromised performances, which is essential to aggregate knowledge to discover unknown attacks.}
\label{figure05}
\end{figure*}

\subsubsection{Datasets and Model Configuration}
Methods for sampling network traffic and syslog data vary with the target objects. The network traffic in the experiment is demonstrated with the pcap data and the csv\_data files, while the syslog data is mainly sampled from audit.log files contained in each ...log/audit folder. As mentioned above, there are three sub-datasets after data pre-processing, one is a dataset of $86,691$ samples containing only traffic data, and the other is a syslog dataset containing only $642$ samples after timestamp alignment, the last one is an extended dataset that assigns the missing syslog data to $0$, which contains $87732$ samples. In the following experiment, the first dataset is used to train a model only with traffic data, the second dataset is used to train another clean model only with syslog data, and the third dataset is used to train an FMKR model.

\subsubsection{Model Settings for Experiments}
The FMKR includes multiple uni-modal models, as mentioned previously, both MLP and ResNet are used to train the multi-modal learning model. For the MLP network of PCAP files, the input layer has 80 neurons; the two hidden layers contain 64 hidden neurons and 32 hidden neurons, respectively. The activation function layer behind each hidden layer adopts the ReLU function. For the MLP network of log files, the input layer has 14 neurons, each of the hidden layers contains 20 neurons, and the activation layer also has 12 neurons. For the ResNet model of PCAP files, there is a $1*3$ two-dimensional convolution layer with $1$ input channel and $2$ output channels, a batch normalization (BN) layer, a $1*3$ global aggregation layer, and then $4$ ResBlock layers. Therein, each ResBlock is composed of $2$ residual modules, the first residual module uses a $1*1$ convolution layer to directly add the input into the output. Moreover, the number of output channels of the next three ResBlocks is twice the number of input channels, and the output shape is half of the input shape. After that, it goes through a $1*3$ global aggregation layer, then uses a flattened layer to reduce the dimension into a one-dimensional vector, and finally passes through 3 fully-connected layers. For the ResNet of log files, there are only two residual blocks. 

\subsection{Evaluation Matrix}
Our experiment analyzes the performance of each model from the following four indicators: model accuracy, model precision, recall rate, and $F_1$-score. Model accuracy is a common and intuitive indicator to measure model performance. However, problems may also arise in the case of imbalanced sample distribution. For example, a testing set contains 95\% of the positive class and the model predicts all inputs as positive classes with 95\% accuracy. Such a model has a very low generalization ability. The model precision is the correct ratio of the classifier on the positive samples, which focuses on whether the classifier is accurate enough on the positive samples. The recall rate is the ratio of the samples of the positive class judged by the classifier to all the samples of the positive class, this indicator focuses on whether the classifier can find more positive samples. 
The $F_1$ score is the harmonic average of model precision and recall rate, which comprehensively measures the classification performance. In our experimental settings, all abnormal samples is labeled as ``positive". The formulas for calculating these indicators are shown as follows:

\begin{equation}
Acc = \frac{TP+TN}{TP+TN+FP+FN}
\end{equation}
\begin{equation}
Prec = \frac{TP}{TP+TN}
\end{equation}
\begin{equation}
RR = \frac{TP}{TP+FN}
\end{equation}
\begin{equation}
F_1 = \frac{2*Prec * RR}{Prec+RR}
\end{equation}


Therein, $TP$ represents the number of samples that are correctly predicted to be abnormal categories. This experimental scene includes the number of correctly-predicted abnormal categories. $TN$ represents the number of samples that are correctly predicted to be normal categories. $FP$ is the number of samples that are incorrectly predicted to be abnormal categories, that is, the number of samples that are actually predicted to be abnormal categories from normal categories, and $FN$ is the number of samples that are incorrectly predicted to be normal categories.

\subsection{Results and Discussion}
In this section, we will introduce experimental results and discuss the performance of the proposed FMKR. Each experiment records the change of the loss function over the 200 epochs and the accuracy of the training and testing sets.

\subsubsection{Baseline Results}
Firstly, we train three models, 1) The traffic-only model with PCAP files, 2) the syslog-only model with csv files; 3) the traffic and syslog-fused model. The results of traffic-only model, syslog-only model, and traffic and syslog-fused model are evaluated using the ResNet. The details are illustrated in Fig. \ref{figure05}. The recall of traffic and syslog-fused models increases rapidly, while the recall of traffic-only models increases slowly. The accuracy of a syslog-only model soared before 150 epochs and then begins to oscillate violently due to a very small number of training samples. Based on these observations, we can find that fusing syslog and network traffic will achieve compromised performances, which is essential to aggregate knowledge to discover unknown attacks. Therefore, we further evaluate the detection ability of the proposed FMKR scheme against SOTA methods.

\begin{table}[ht]
    \centering
    \footnotesize
    \renewcommand\arraystretch{1.3}
    \caption{Comparison results of different few-shot threat detection methods against APT attacks}
    \begin{tabular}{c|cc|cc|cc}
\toprule
Attack & \multicolumn{2}{c|} {FC-Net} & \multicolumn{2}{c|} {FMKR} & \multicolumn{2}{c} {Fine-tuned FMKR}\\
stages & ACC  & F1 & ACC & F1 & ACC & F1 \\
\hline
NT & 97.99 & 96.05 & 97.55 & 95.22 & 97.82 & 95.73 \\
RN & 96.10 & 92.49 & 95.49 & 91.37 & 96.20 & 92.31 \\
EF & 95.32 & 91.06 & 94.46 & 89.50 & 96.11 & 92.51 \\
LM & 70.04 & 53.87 & 93.12 & 87.16 & 93.98 & 88.64 \\
DE & \textbf{16.67} & \textbf{9.09} & \textbf{83.54} & \textbf{71.73} & 100.0 & 100.0\\
Total& 96.66 & 93.54 & 96.15 & 92.59 & \textbf{97.33} & \textbf{94.8}\\
\bottomrule
    \end{tabular}
    \label{tab:results_cifar_other_classes}
\end{table}

\subsubsection{Comparisons to SOTA Methods}
Few-shot threat detection is a relatively emerging topic so there are only a few related studies available for comparisons. Therefore, to further verify the superiority of the proposed FMKR, we modified the FC-Net settings in \cite{9083983} to complete the threat detection task based on DAPT 2020 dataset. The detection results are compared in Table I. It can be seen that the FMKR performs better F1 score than existing methods on the samples of the ``DE" stage. Therein, ``NT" stands for ``normal traffic" in the dataset, ``RN" is the ``Reconnaissance" stage, ``EF" denotes ``Establish Foothold", ``LM" represents for ``Lateral Movement", and ``DE" is the ``Data Exfiltration".

\subsection{Ablation Experiments of Proposed FMKR}
\subsubsection{Reasonable batch size is important for the FMKR convergence.} The suitable hyper-parameters for different ResNet-based FMKR models are not consistent, and the actual training should also depend on the specific situation. To present the impact factors to the performance of proposed FMKR scheme, we evaluate the FMKR under different batch sizes: 128, 64, and 32, respectively. The FMKR model is more suitable for the hyper-parameters with a batch size of 128, as illustrated in Fig. 5. The accuracy of the ResNet-based FMKR model fluctuates violently when the batch size is 32 or 128. When the batch size is 64, the accuracy changes smoothly and other performance indicators are also better, indicating that the batch size of 64 is the most suitable for ResNet-based FMKR.

\begin{figure}[ht]
\centering
\includegraphics[width=3.5in]{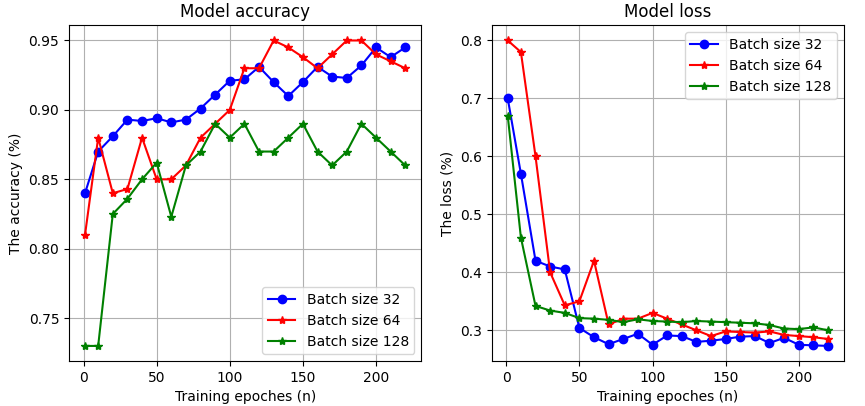}
\caption{The FMKR performance under different batch sizes.}
\label{figure06}
\end{figure}

\subsubsection{The defense satisfaction rate is sensitive to the number of fresh samples for model fine-tuning.}
As introduced in prevision, the FMKR is cloud-edge synergistic. Before being deployed on endpoints, the FMKR models should be fine-tuned with fresh samples. Model fine-tuning can be initiated through endpoint requests or periodically updated through the cloud. To maximize available knowledge of multiple network domains, cloud data center can collect model parameters from distributed endpoints and aggregate them as a global model. After aggregating, endpoints can fine-tune the global model. In our experiments, we consider four different strategies: 

\begin{itemize}
    \item Highest accuracy first, the operator monitors all fine-tuned FMKR models on edges/endpoints, and select the model with highest accuracy as the fine-tuned model in the next training round. 
    \item Fine-tune after aggregation, introducing the aggregation algorithm of federated learning to achieve a unified model at cloud data center, and distribute the aggregated model to each endpoint for fine-tuning.
    \item Random fine-tuning, randomly select a model from the endpoints as the fine-tuned model in the next training round.
    \item Fine-tuning itself. Take itself as the model to fine-tune in the next training round.
\end{itemize}
\begin{figure}[ht]
\centering
\includegraphics[width=3.0in]{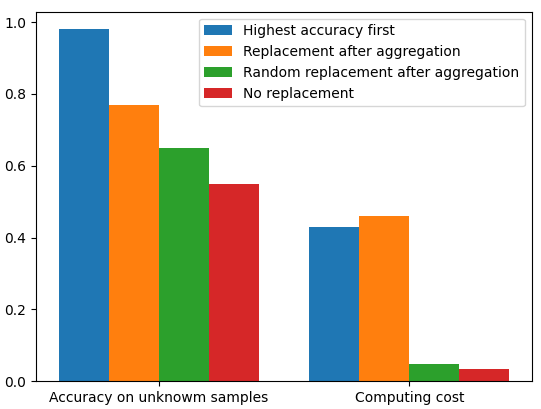}
\caption{Observations results under different model deployment strategies. Each strategy works 15 days.}
\label{figure06}
\end{figure}

Fig. 6 shows the model accuracy on unknown attack samples and computing cost of proposed FMKR under different model deployment strategies. Since both ``highest accuracy first" and ``fine-tune after aggregation" strategies require to retrieve all edges/endpoints, the computing cost will increase with the number of participants. In our observations, the number of participants is configured as ``10", thus their computing costs are about 10 times of ``random replacement" and ``fine-tuning itself". 

\section{Conclusion}
In this paper, a few-shot multi-domain knowledge rearming (FMKR) scehme is proposed for context-aware APT detection. By completing multiple small tasks that are generated from different network domains with meta-learning, the FMKR firstly trains a model with good discrimination and generalization ability for fresh and unknown APT attacks. In each FMKR task, both threat intelligence and local entities are fused into the support/query sets in the meta-learning task to identify possible attack stages. Secondly, to rearm current security strategies, an finetuning-based deployment mechanism is also proposed to transfer learned knowledge into the local model, while minimizing the defense cost. Compared to multiple model replacement strategies, the FMKR provides a faster response to attack behaviors while consuming less computing cost. Based on the feedback from multiple real IIoT users over 2 months, we demonstrate that the proposed scheme can improve the defense satisfaction rate.


\section{Acknowledgments}
This research work is funded by the National Nature Science Foundation of China under Grant No. 62202303 and U20B2048, the Fundamental Research Funds for the Central Universities, and Shanghai Sailing Program under Grant No. 21YF1421700.

\balance
\bibliographystyle{ieeetr}
\bibliography{ref}
\end{document}